\documentstyle[]{mn}
\newif\ifAMStwofonts

\def\etal{{\rm et al.}}

\def\simgt{\mathrel{\spose{\lower 3pt\hbox{$\sim$}}
        \raise 2.0pt\hbox{$>$}}}
\def\simlt{\mathrel{\spose{\lower 3pt\hbox{$\sim$}}
        \raise 2.0pt\hbox{$<$}}}

\ifoldfss
  \newcommand{\rmn}[1] {{\rm #1}}

  \ifCUPmtlplainloaded \else
    \NewTextAlphabet{textbfit} {cmbxti10} {}
    \NewTextAlphabet{textbfss} {cmssbx10} {}
    \NewMathAlphabet{mathbfit} {cmbxti10} {} 
    \NewMathAlphabet{mathbfss} {cmssbx10} {} 
  \fi
  \ifAMStwofonts
    \ifCUPmtlplainloaded \else
      \NewSymbolFont{upmath} {eurm10}
      \NewSymbolFont{AMSa} {msam10}
      \NewMathSymbol{\upi}     {0}{upmath}{19}
      \NewMathSymbol{\umu}     {0}{upmath}{16}
      \NewMathSymbol{\upartial}{0}{upmath}{40}
      \NewMathSymbol{\leqslant}{3}{AMSa}{36}
      \NewMathSymbol{\geqslant}{3}{AMSa}{3E}

    \fi
  \fi
\fi 

\ifnfssone
  \newmathalphabet{\mathit}
  \addtoversion{normal}{\mathit}{cmr}{m}{it}
  \addtoversion{bold}{\mathit}{cmr}{bx}{it}
  \newcommand{\rmn}[1] {\mathrm{#1}}

  \newmathalphabet{\mathbfit} 
  \addtoversion{normal}{\mathbfit}{cmr}{bx}{it}
  \addtoversion{bold}{\mathbfit}{cmr}{bx}{it}
  \newmathalphabet{\mathbfss} 
  \addtoversion{normal}{\mathbfss}{cmss}{bx}{n}
  \addtoversion{bold}{\mathbfss}{cmss}{bx}{n}
  \ifAMStwofonts
    \ifCUPmtlplainloaded \else
      \UseAMStwoboldmath
      \makeatletter
      \new@mathgroup\upmath@group
      \define@mathgroup\mv@normal\upmath@group{eur}{m}{n}
      \define@mathgroup\mv@bold\upmath@group{eur}{b}{n}
      \edef\UPM{\hexnumber\upmath@group}
      \new@mathgroup\amsa@group
      \define@mathgroup\mv@normal\amsa@group{msa}{m}{n}
      \define@mathgroup\mv@bold\amsa@group{msa}{m}{n}
      \edef\AMSa{\hexnumber\amsa@group}
      \makeatother
      \mathchardef\upi="0\UPM19
      \mathchardef\umu="0\UPM16
      \mathchardef\upartial="0\UPM40
      \mathchardef\leqslant="3\AMSa36
      \mathchardef\geqslant="3\AMSa3E
    \fi
  \fi
\fi 

\ifnfsstwo
  \newcommand{\rmn}[1] {\mathrm{#1}}

  \DeclareMathAlphabet{\mathbfit}{OT1}{cmr}{bx}{it}
  \SetMathAlphabet\mathbfit{bold}{OT1}{cmr}{bx}{it}
  \DeclareMathAlphabet{\mathbfss}{OT1}{cmss}{bx}{n}
  \SetMathAlphabet\mathbfss{bold}{OT1}{cmss}{bx}{n}
  \ifAMStwofonts
    \ifCUPmtlplainloaded \else
      \DeclareSymbolFont{UPM}{U}{eur}{m}{n}
      \SetSymbolFont{UPM}{bold}{U}{eur}{b}{n}
      \DeclareSymbolFont{AMSa}{U}{msa}{m}{n}
      \DeclareMathSymbol{\upi}{0}{UPM}{"19}
      \DeclareMathSymbol{\umu}{0}{UPM}{"16}
      \DeclareMathSymbol{\upartial}{0}{UPM}{"40}
      \DeclareMathSymbol{\leqslant}{3}{AMSa}{"36}
      \DeclareMathSymbol{\geqslant}{3}{AMSa}{"3E}
    \fi
  \fi
\fi 

\ifCUPmtlplainloaded \else
  \ifAMStwofonts \else 
    \def\upi{\pi}
    \def\umu{\mu}
    \def\upartial{\partial}
  \fi
\fi

\title[The distributions of microlensed light curve derivatives]
  {The distribution of microlensed light curve derivatives:
 the relationship between stellar proper motions and transverse velocity}
\author[J. S. B. Wyithe, R. L. Webster \& E. L. Turner]
  {J.~S.~B.~Wyithe$^{1,2}$, 
  R.~L.~Webster$^1$,
  E. L.~Turner$^2$,\\
  $^1$ School of Physics, University of Melbourne, Parkville, Vic, 3052, 
Australia\\
  $^2$ Princeton University Observatory, Peyton Hall, Princeton, NJ 08544 USA\\ 
 Email: swyithe@physics.unimelb.edu.au, rwebster@physics.unimelb.edu.au, elt@astro.princeton.edu }
\date{Accepted 11. Received }
\pagerange{\pageref{firstpage}--\pageref{lastpage}}
\pubyear{1998}

\def\LaTeX{L\kern-.36em\raise.3ex\hbox{a}\kern-.15em
    T\kern-.1667em\lower.7ex\hbox{E}\kern-.125emX}

\begin{document}

\label{firstpage}

\maketitle

\begin{abstract}

We present a method for computing the probability distribution of microlensed light curve derivatives both in the case of a static lens with a transverse velocity, and in the case of microlensing that is produced through stellar proper motions. The distributions are closely related in form, and can be considered equivalent after appropriate scaling of the input transverse velocity. The comparison of the distributions in this manner provides a consistent way to consider the relative contribution to microlensing (both large and small fluctuations) of the two classes of motion, a problem that is otherwise an extremely expensive computational exercise. We find that the relative contribution of stellar proper motions to the microlensing rate is independent of the mass function assumed for the microlenses, but is a function of optical depth and shear. We find that stellar proper motions produce a higher overall microlensing rate than a transverse velocity of the same magnitude. This effect becomes more pronounced at higher optical depth. With the introduction of shear, the relative rates of microlensing become dependent on the direction of the transverse velocity. This may have important consequences in the case of quadruply lensed quasars such as Q2237+0305, where the alignment of the shear vector with the source trajectory varies between images.
\end{abstract}


\section{Introduction}

Calculations of the properties of a light curve from a gravitationally microlensed quasar have traditionally assumed that the variation in observed flux is the result of an effective transverse velocity of the source with respect to the foreground galactic starfield. These calculations have assumed that the contribution to microlensing from the random proper motions of stars is a negligible one because the galactic transverse velocity is an order of magnitude larger than the typical stellar proper motion. However in the case of Q2237+0305, the only object in which cosmological microlensing has been confirmed (Irwin et al. 1989; Corrigan et al. 1991), the line-of-sight velocity dispersion is comparable to the likely transverse velocity. Foltz et al. (1992) measured the central velocity dispersion of Q2237+0305 to be $\sim$ 215$\,km\,sec^{-1}$. This is to be compared with the estimated galactic transverse velocity of $\sim$ 600$\,km\,sec^{-1}$, and the suggestion by Witt \& Mao (1993) who provide evidence from the observed high magnification events (HMEs) for a smaller transverse velocity. Q2237+0305 is the perfect object from which to study microlensing.  The source quasar is lensed by an intervening galaxy, producing 4 images, each of which is observed through the galactic bulge with an optical depth in stars that is of order unity (eg. Schmidt, Webster \& Lewis 1998). In addition, the close proximity of the lensing galaxy means that the effective transverse velocity may be high.
 However the potential contribution of the velocity dispersion makes accurate direct modelling of the microlensing light curves difficult. Therefore while Q2237+0305 is the best object from which to observe microlensing, a realistic numerical analysis of the observations is difficult. 

This paper describes the relative contributions to microlensing of random stellar proper motions, and of a transverse velocity of the entire lens. Two approaches have been used to tackle this problem in the past. Kundic \& Wambsganss (1993), and Wambsganss \& Kundic (1995) computed light curves that included the effect of stellar proper motions by producing a magnification map at many successive intervals for an evolving field of point masses. Light curves that resulted from the random motion of stars alone were then produced by looking at the changing magnification at a fixed source point, and compared with light curves that resulted from a transverse velocity of the source with respect to one of the starfields. These light curves were produced in the usual way by shifting the magnification pattern about the source point. Schramm et al. (1992) obtained an expression for the caustic velocities that result from a stream motion of the star field. This idea was extended by Kundic, Witt \& Chang (1993) who calculated the caustic velocity resulting from a stellar velocity dispersion. They calculated the area swept out per unit time by the caustic network in both the cases of a lens composed of a static field of point masses with a transverse velocity, and one where the the point masses are given a random proper motion. These papers describe the effect that proper motions have on the statistics of high magnification events (HMEs). Unfortunately their conclusions are not consistent. Kundic, Witt \& Chang (1993) claim that the relative efficiency of stellar proper motion and transverse velocity in terms of the resulting frequency of HMEs is a function of the optical depth, while Kundic \& Wambsganss (1993) find no such dependence. The effect of the inclusion of a shear in the models is mentioned in these papers. However no conclusions are presented regarding variation in the relative importance of microlensing due to stellar proper motions with the direction of the transverse velocity. 

Our approach is to compute the distribution of light curve derivatives in the cases of a static lens with a transverse velocity, and of a static source point in a model where the point masses are given a random proper motion. As shown below these distributions are very similar up to a scaling factor in the derivative and can be quantitatively compared in order to find the relative contributions to microlensing of the two cases. The advantages of this approach are two fold. Firstly, the average distribution of light-curve derivatives can be computed from points that are not sequential along a light curve. {\em The distribution of light-curve derivatives for the cases of proper motion and transverse velocity can therefore be computed at the same computational expense}. This is in contrast to the direct computation of a light curve (Kundic \& Wambsganss 1993; Wambsganss \& Kundic 1995) where the increase in time is equal to the number of points on the light curve.
 Secondly, the distribution of light-curve derivatives provides information on the microlensing rate in all derivative regimes. It is therefore more suited to the analysis of monitoring data than a method that considers only the frequency of caustic crossings because HMEs are rare, and may be missed by the low sampling rate. 

 This paper is concerned with the construction of the histogram, the nature of microlensing due to stellar proper motions, as well as its dependence on optical depth, shear and direction of the source trajectory. It is presented in 8 parts. Sections \ref{applic} and \ref{models} discuss the utility of the distribution of microlensing light-curve derivatives, and describe the microlensing models used in the paper. Sections \ref{const} and \ref{comp} describe the construction of the derivative histograms for the proper motion and transverse velocity cases, and the method for their comparison. In Sections \ref{opt_sect} and \ref{shear_sect} the variation with optical depth and shear of the relative contribution of proper motions and transverse velocity to the microlensing rate are discussed. Section \ref{2237_sect} presents the relative rates for the parameters of optical depth and shear corresponding to the images of Q2237+0305.

\section{Application of the distribution of Microlensed light-curve Derivatives} \label{applic} 

 This paper is part of an analysis of the 12 years of published monitoring data for Q2237+0305. In this analysis the distribution of light curve derivatives is considered, an approach that yields information on both the effective galactic transverse velocity, the source size and the microlens mass function. These aspects are discussed in subsequent papers (Wyithe, Webster \& Turner 1999b,c; Wyithe, Webster, Turner \& Mortlock 1999), but the distribution of the derivatives due to proper motion of stars is an integral part of the analysis. The gain in computational speed makes consideration of proper motions in a large range of models feasible. However, the histogram of microlensed light curve derivatives does not contain information on the structure of the light-curve. 
 Therefore, to apply the derivative distributions due to proper motions to a set of monitoring data we make the assumption that a typical section of light-curve resulting from stellar proper motions will have similar statistics to a section of light curve produced by a static lens in combination with an appropriate transverse velocity. Furthermore, for the distribution of derivatives to be applicable to short sections of light-curve, caustic crossings in the two cases should have the same clustering properties.

 To justify these assertions we note firstly that (as shown below) at an appropriate transverse velocity, light curves from a static field have a very similar average distribution of microlensed light-curve derivatives to that produced by proper motions, indicating equal average rates of light curve peaks and average peak durations (the peak heights must also be equal). In addition, the similarity of the two classes of derivative distribution suggests an approximately equal temporal clustering of peaks since additional clustering in one case would produce an excess of small derivatives and a corresponding dearth of mid-range derivatives over the other case. Secondly, at any given time caustics that move under the influence of random proper motions have the same spatial clustering since there is no correlation between motions of individual point masses. Also, in general the motion of two caustics that are in close proximity are governed primarily by the motions of separate stars or groups of stars. Caustics that have spatial correlation therefore do not have a corresponding correlation in the direction or magnitude of their velocity, and hence there should be no additional temporal correlation of caustic crossings.

 With these points in mind, we subsequently assume that the temporal clustering of caustic crossings is the same for the two classes of motion, and that light-curves produced by proper motion and by an appropriate transverse velocity can be considered equivalent. Sampled model light-curves due to stellar proper motions can then be produced from static model light curves with the appropriate transverse velocity and compared with observation.

\section{The microlensing models}
\label{models}
\begin{table*}
\caption{\label{parameters}The parameters of microlensing models considered in this paper.}
\begin{tabular}{|c|c|c|c|c|c|c|c|c|}
\hline
simulation no.   & $\kappa$  &  $\vec{\gamma}=(\gamma_{1},\gamma_{2})$ &  simulation type   &  source line length  &  no. stars  &  no. simulations  & $\langle \mu_{av} \rangle$ & $\langle \mu_{th} \rangle$ \\ \hline\hline
1                 &  0.025    & (0,0)     &      $MSOLAR$     &  50$\eta_{o}$        & 500         & 500               & 1.04$\pm$.01               &  1.05                          \\
2                 &  0.05     &  (0,0)    &      $MSOLAR$     &  50$\eta_{o}$        & 500         & 500               & 1.08$\pm$.01              &  1.11                          \\
3                 &  0.1      &  (0,0)    &      $MSOLAR$     &   50$\eta_{o}$       & 500         & 500               & 1.18$\pm$.01              &  1.23                          \\
4                 &  0.2      &  (0,0)    &       $MSOLAR$    &   50$\eta_{o}$       & 511         & 250               & 1.54$\pm$.04              &  1.56                          \\
5                 &  0.3      &  (0,0)    &       $MSOLAR$    &    50$\eta_{o}$      & 1127        & 250               & 2.02$\pm$.05              &  2.04                          \\
6                 &  0.4      &  (0,0)    &       $MSOLAR$    &  25$\eta_{o}$        & 1071        & 250               & 2.68$\pm$.05              &  2.78                          \\
7                 &  0.5      &  (0,0)    &       $MSOLAR$    &      25$\eta_{o}$    & 2194        & 250               & 4.09$\pm$.10              &  4.00                          \\
8                 &  0.6      &  (0,0)    &       $MSOLAR$    &      25$\eta_{o}$    & 4597        & 100               & 6.16$\pm$.15              &  6.25                          \\
9                 &  0.2      &  (0,0)    &       $SALPETER$  &     50$\eta_{o}$     & 1792        & 200               & 1.51$\pm$.02             &  1.56                          \\
10                &  0.3      &  (0,0)    &       $SALPETER$  &     50$\eta_{o}$     & 3792        & 200               & 2.03$\pm$.05              & 2.04                           \\
11                &  0.4      &  (0,0)    &       $SALPETER$  &      50$\eta_{o}$    & 7333        & 200               &  2.63$\pm$.05              & 2.78                           \\
12                &  0.5      &  (0,0)    &       $SALPETER$  &       25$\eta_{o}$   & 5611        & 100               &  3.95$\pm$.12             & 4.00                           \\
13                &  0.6      &  (0,0)    &       $SALPETER$  &      10$\eta_{o}$    & 5423        & 100               & 6.26 $\pm$.33             & 6.25                           \\
14                &  0.3      &  (-0.4,0) &       $MSOLAR$    &     50$\eta_{o}$     & 855         & 200               & 2.95$\pm$.09              & 3.03                           \\
15                &  0.3      &  (-0.2,0) &       $MSOLAR$    &     50$\eta_{o}$     & 789         & 200               & 2.18$\pm$.04              & 2.22                           \\
16                &  0.3      &   (0.2,0) &       $MSOLAR$    &     50$\eta_{o}$     & 2102        & 200               & 2.23$\pm$.03             & 2.22                           \\
17                &  0.3      &   (0.4,0) &        $MSOLAR$   &     50$\eta_{o}$     & 5739        & 200               & 3.00$\pm$.01              & 3.03                           \\
18     		  &0.36       &   (0.4,0) &        $SALPETER$ &     10$\eta_{o}$     & 2926        & 60                & 3.92$\pm$.08                  & 4.01                           \\
19     		  &0.36       &(0.28,0.28)&        $SALPETER$ &     10$\eta_{o}$     & 1774        & 60                & 3.95$\pm$.14                 & 4.01		              \\
20     		  &0.36       &   (-0.4,0) &        $SALPETER$ &     10$\eta_{o}$     & 643         & 60                & 3.75$\pm$.15                  & 4.01		             \\
21     		  &0.69       &   (0.71,0)&        $SALPETER$ &     10$\eta_{o}$     &  2782       & 30                &    2.52$\pm$.17                 & 2.45		             \\
22     		  &0.69       &(0.50,0.50)&        $SALPETER$ &     10$\eta_{o}$     &  2311       & 30                & 2.46$\pm$.10                 & 2.45		             \\ 
23     		  &0.69       &   (-0.71,0)&        $SALPETER$ &     10$\eta_{o}$     &  1058       & 30                &    2.42$\pm$.10                  & 2.45		             \\
24     		  &0.59       &   (0.61,0)&        $SALPETER$ &     10$\eta_{o}$     & 8583        & 30                &     4.77$\pm$.10                 & 4.90		             \\
25     		  &0.59       &(0.43,0.43)&        $SALPETER$ &     10$\eta_{o}$     & 6675        & 30                &     4.82$\pm$.29                 & 4.90		             \\
26     		  &0.59       &   (-0.61,0)&        $SALPETER$ &     10$\eta_{o}$     & 2219        & 30                &    4.62$\pm$.44                  & 4.90		            \\ \hline
\end{tabular}

\end{table*}

 Throughout the paper, standard notation for gravitational lensing is used. The Einstein radius of a 1$M_{\odot} $ star in the source and image planes are denoted by $\eta_{o}$ and $\xi_{o}$ respectively. The normalised shear due to external mass is denoted by $\vec{\gamma}=(\gamma_{1},\gamma_{2})$, and the convergence or optical depth by $\kappa$. The model for gravitational microlensing consists of a very large sheet of point masses that simulates the section of galaxy along the image line of sight, together with a shear term that includes the perturbing effect of the mass distribution of the lensing galaxy as a whole. The normalised lens equation for a field of point masses with an applied shear in terms of these quantities is
\begin{equation}
\vec{y}= \left( \begin{array}{cc}
	 1-\gamma_{1} & -\gamma_{2} \\
	-\gamma_{2} & 1+\gamma_{1} 
	    \end{array} \right)\vec{x} + \sum_{j=0}^{N_{*}}m^{j}\frac{(\vec{x}^{j}-\vec{x})}{|\vec{x}^{j}-\vec{x}|^{2}}
\label{lens_map} 
\end{equation}
Here $\vec{x}$ and $\vec{y}$ are the normalised image and source positions respectively, and the $\vec{x_{i}}$ are the normalised positions of the point masses. The magnification, or increase in flux due to gravitational lensing of a point source is obtained from the inverse of the Jacobian determinant of Eqn~\ref{lens_map}:

\begin{equation}
\mu_{p}=\frac{1}{det(A(\vec{x}))},
\end{equation}

\noindent where $A\equiv\frac{\partial \vec{y}(\vec{x})}{\partial\vec{x}}$ is the Jacobian matrix. The magnification $\mu_{p}$ may be negative, corresponding to the magnification of an inverted image. In the case of multiple imaging of a point source the magnification $\mu_{p}$ is given by

\begin{equation}
\mu_{p}=\sum_{images}|\mu_{images}|.
\label{mu_im}
\end{equation}

We consider two classes of model in our analysis, both consisting of point masses with positions that are distributed randomly in a disc. In sections \ref{opt_sect} and \ref{shear_sect} each model star is assigned a mass of 1$M_{\odot}$. These simulations are termed to be of type $MSOLAR$. In sections \ref{mass_func_sect}, \ref{vel_disp_sect} and \ref{2237_sect} the model is based on that of Wambsganss, Paczynski \& Katz (1989), in which masses are distributed according to a Salpeter mass function ($p(m)dm\propto m^{-2.35}$) in the range $0.1M_{\odot}<m_{i}<1.0M_{\odot}$. The label $SALPETER$ refers to simulations of this type. None of the models presented in this paper include a component of continuously distributed matter $\kappa_{c}$ in addition to an optical depth in stars $\kappa_{*}$. The results can however be applied to this case since a model that includes continuous matter is mathematically equivalent (through the parameter transformation of Paczynski (1986)) to a model that contains only compact objects, but which is described by a different set of microlensing parameters ($\kappa^{\prime}$ and $\gamma^{\prime}$) and different scale lengths ($\xi^{\prime}_{o}$ and $\eta^{\prime}_{o}$). For the case where $\kappa_{c}<1$:
\begin{eqnarray}
\nonumber
\kappa^{\prime}&=&\frac{\kappa_{*}}{1-\kappa_{c}}\\\nonumber
\gamma^{\prime}&=&\gamma\times(1-\kappa_{c})\\\nonumber
\xi^{\prime}_{o}&=&\frac{\xi_{o}}{\sqrt{1-\kappa_{c}}}\\\
\eta^{\prime}_{o}&=&\eta_{o}\times\sqrt{1-\kappa_{c}}
\end{eqnarray}
The point source magnifications are related by 
\begin{eqnarray}
\mu^{\prime}=(1-\kappa_{c})^{2}\mu.
\end{eqnarray}

\begin{figure}
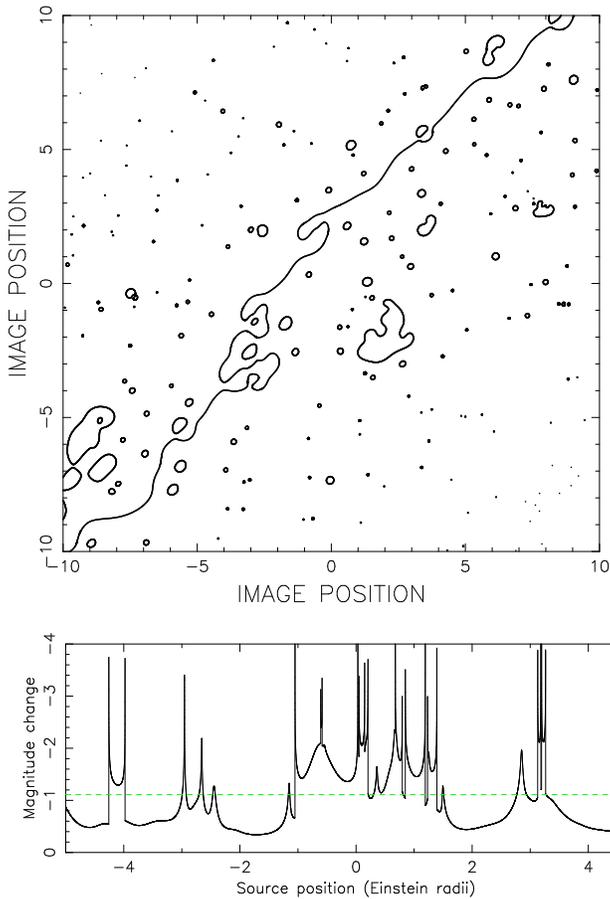

\vspace*{120mm}
\centerline {
\includegraphics{fig1a.epsi}
\includegraphics{fig1b.epsi}
}
\caption{ The images of a source line (top) and point source light curve (bottom) for a simulation having an optical depth of $\kappa=0.4$. The source trajectory is aligned at 45 degrees to the shear vector. The dashed light line shows the theoretical average magnification for this model.}
\label{imlc}
\end{figure}

 The region of the lens plane in which image solutions need to be found to ensure that $99\%$ of the total macro-image flux is recovered from a source point was described by Katz, Balbus \& Paczynski (1986). In the presence of an applied shear these regions are elliptical in shape. The union of the areas of the lens plane that correspond to the flux collection area of each point on the source line is known as the shooting region. The method of determining the dimensions of the shooting region is described in Lewis \& Irwin (1995), and Wyithe \& Webster (1998). The radius of the disc of point masses is chosen to be $1.2\times$ that required to cover this shooting region. Eqn \ref{lens_map} is solved through the inversion technique of Lewis et al. (1993) and Witt (1993). This technique finds all image solutions of a source line that lie within the shooting region using the fact that these lie on one of two classes of image curve. The first of these curves is a line that tends asymptotically to the unperturbed (by the point masses) image of the source line far from the starfields edge. This line may pass over one or more of the stars. The second class of image curve is a collection of loops, each of which passes over one or more stars. An example of the lensed images of a line is shown in Figure \ref{imlc}. We implement the contouring method as described by Lewis et al. (1993). Image solutions are found iteratively along the image curves. When each new image point is located the image magnifications are added to predefined source points through an interpolation between the current and previous steps. The interpolation schemes between two source points ($i$ and $i+1$) were given by Lewis et al. (1993):

\begin{equation}
y \propto \mu_{image}^{\alpha}, \\
\end{equation}
where 
\begin{eqnarray}
\alpha&=&-2\,\,\,\,{\rmn for}\,\,\, \mu_{i}\,\,\,{\rmn or}\,\,\,\mu_{i+1}>\frac{5}{|(1-\kappa)^{2}-\gamma^{2}|},  \\ 
\alpha&=&-\frac{1}{4}\,\,\,\,{\rmn for}\,\,\, \mu_{i}\,\,\,{\rmn and}\,\,\,\mu_{i+1}<0.2, \\
\alpha&=&1\,\,\,\,{\rmn otherwise}.
\end{eqnarray}

Table \ref{parameters} displays the parameters of simulations discussed in this paper. Each simulation is labelled numerically, and is described by the associated values of $\kappa$ and $\vec{\gamma}$, as well as the simulation type. Columns 5 and 6 in Table \ref{parameters} show the length of source track and the number of stars in each individual simulation respectively. Column 7 shows the number of individual simulations in the set. The final 2 columns show the mean amplification resulting from the set of simulations, and the theoretical value ($\mu_{th}=|(1-\kappa)^{2}-|\vec{\gamma}|^{2}|^{-1}$) for comparison.
Each set of simulations was divided into 5 subsets, and the error in the mean magnification of the models estimated from the standard deviation in the resulting values.
 Figure \ref{amp_dist} demonstrates that the magnification distributions of our simulations are independent of mass function and source trajectory direction as required.

\section{Construction of the Histogram of Derivatives}
\label{const}

To calculate the distribution of light curve derivatives that result from stellar proper motions, the derivative of the magnification is taken with respect to time at the position of each image. These derivatives are then added to give the total change in magnification due to the movement of the point masses. Eqn \ref{lens_map} can be written in component form:
\begin{equation}
y_{1}=(1-\gamma)x_{1}-\sum_{j=0}^{N_{*}}m^{j}\frac{x_{1}-x_{1}^{j}}{(x_{1}-x_{1}^{j})^{2}+(x_{2}-x_{2}^{j})^{2}}
\end{equation}
and
\begin{equation}
y_{2}=(1+\gamma)x_{2}-\sum_{j=0}^{N_{*}}m^{j}\frac{x_{2}-x_{2}^{j}}{(x_{1}-x_{1}^{j})^{2}+(x_{2}-x_{2}^{j})^{2}},
\end{equation}
where $x_{i}^{j},m^{j},x_{i},y_{i}$, for $i=1,2$ are the position and mass of the $j$th star, and the position of the image and source respectively. For simplicity the source and image axes have been aligned with the shear vector in this expression $\left(\vec{\gamma}\equiv(\gamma,0)\right)$ in order to diagonalise the shear matrix.
 The positions of the point masses are functions of time:
\begin{equation}
v^{j}_{i}(t)\equiv\frac{d\,x_{i}^{j}}{d\,t}.
\end{equation}
For a given image we write down the expression for the derivative of magnification with respect to time in terms of these velocities: 
\begin{eqnarray}
\nonumber
\left.\frac{d\,\mu_{p}}{d\,t}\right)_{proper}&=&\frac{\partial\,\mu_{p}}{\partial\,x_{1}}\left( \sum_{j=0}^{N_{*}} \frac{\partial\,x_{1}}{\partial\,x_{1}^{j}}\frac{d\,x_{1}^{j}}{d\,t} + \sum_{j=0}^{N_{*}} \frac{\partial\,x_{1}}{\partial\,x_{2}^{j}}\frac{d\,x_{2}^{j}}{d\,t} \right) \\ \nonumber
                   &+&\frac{\partial\,\mu_{p}}{\partial\,x_{2}}\left( \sum_{j=0}^{N_{*}} \frac{\partial\,x_{2}}{\partial\,x_{1}^{j}}\frac{d\,x_{1}^{j}}{d\,t} + \sum_{j=0}^{N_{*}} \frac{\partial\,x_{2}}{\partial\,x_{2}^{j}}\frac{d\,x_{2}^{j}}{d\,t} \right) \\
                   &+& \sum_{j=0}^{N_{*}} \frac{\partial\,\mu_{p}}{\partial\,x_{1}^{j}}\frac{\partial\,x_{1}^{j}}{\partial\,t} + \sum_{j=0}^{N_{*}} \frac{\partial\,\mu_{p}}{\partial\,x_{2}^{j}}\frac{d\,x_{2}^{j}}{d\,t}
\end{eqnarray}
\begin{figure}
\vspace*{60mm}
\centerline {
\includegraphics{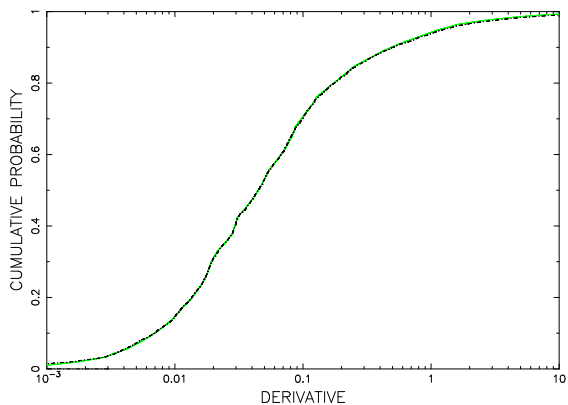}
}
\caption{The cumulative histogram of derivatives (in units of magnitudes per year) calculated from the light curve in Figure \ref{imlc} (dot-dashed line). Each of the point masses in this simulation were given a velocity parallel, opposite in direction, and of the same magnitude as the transverse velocity. The distribution of derivatives was then calculated for this same simulation using the expression \ref{der}, and is shown as the solid light line}
\label{dertest}
\end{figure}

\noindent To compute the derivatives $\frac{\partial\,x_{i}}{\partial\,x_{k}^{j}}$ for $i,k=1,2$ the derivatives $\frac{d\,y_{1}}{d\,x_{k}^i}=0$ and $\frac{d\,y_{2}}{d\,x_{k}^i}=0$ are taken simultaneously. The full expression for $\frac{d\,\mu_{p}}{d\,t})_{proper}$ is given in appendix \ref{app1} (Eqn \ref{der}).

As with the magnifications, the light curve derivatives must be interpolated between iterations onto predefined source points. Where the magnification derivative is the result of stellar proper motions, the interpolation is found by noting that if $y\propto \mu_{p}^{\alpha}$ then $y\propto (\mu_{p}+\frac{d\,\mu_{p}}{d\,t}\Delta t)^{\alpha}$ also holds for small $\Delta t$. Following this we obtain an interpolation scheme for the derivative: 
\begin{equation}
y\propto \mu_{p}^{\alpha-1}\frac{d\,\mu_{p}}{d\,t}.
\label{interp}
\end{equation} 
From Eqn \ref{mu_im}, the time derivative of the point source magnification is
\begin{equation}
\frac{d\,\mu_{p}}{d\,t}=\sum_{i=0}^{N_{*}}sign(\mu_{i})\frac{d\,\mu_{i}}{d\,t}.
\label{mag}
\end{equation}
The magnification is not directly measurable, however we can find the magnification derivative in terms of the resultant rate of change of magnitude.
\begin{equation}
\frac{d\,\Delta m}{d\,t}=\frac{1.09}{\mu_{p}}\frac{d\,\mu_{p}}{d\,t}
\label{mag_der}
\end{equation}

The histogram of derivatives for a static microlensing model can be calculated in a straight forward manner directly from the combination of a computed light curve with a transverse velocity of the lens ($\vec{v}_{tran}=\frac{d\,\vec{y}}{d\,t}$). However, in order to obtain a derivative of comparable accuracy to that calculated from Eqn \ref{der}, we find the rate of change in magnification in this case analytically as before. 
\begin{equation}
\left.\frac{d\,\mu_{p}}{d\,t}\right)_{tran}=\left(\frac{\partial\,\mu_{p}}{\partial\,x_{1}}\frac{\partial\,x_{1}}{\partial\,y_{1}}+\frac{\partial\,\mu_{p}}{\partial\,x_{2}}\frac{\partial\,x_{2}}{\partial\,y_{1}}\right)\frac{d\,y_{1}}{d\,t}
\label{static}
\end{equation}
In this expression $\frac{\partial\,x_{1}}{\partial\,y_{1}}$, and $\frac{\partial\,x_{2}}{\partial\,y_{1}}$ are evaluated by taking the derivatives $\frac{d\,y_{1}}{d\,y_{1}}=1$  and $\frac{d\,y_{2}}{d\,y_{1}}=m$ simultaneously where the source track is described by $y_{2}=m\,y_{1}$. The full expression for $\frac{d\,\mu_{p}}{d\,t})_{tran}$ is given in appendix \ref{app1} (Eqn \ref{der_stat}). The interpolation scheme in this case is simply the time derivative of that for the computed light curve and so Eqn \ref{interp} also holds in this case. In addition, Eqns \ref{mag} and \ref{mag_der} are correct in this case.

Kundic, Witt \& Chang (1993) describe the relationship between the caustic velocity that results from a galactic transverse velocity ($v_{tran}$), and that resulting from a stream velocity of the local starfield ($v_{stream}$). The microlensed light curve derivative is proportional to the caustic velocity, and so we find for a static starfield having both a transverse and a stream motion that 

\begin{equation}
\frac{d\,\mu_{p}}{d\,t} \propto \left| \vec{v}_{tran} + \left( \begin{array}{cc}
	 1-\gamma & 0 \\
	0 & 1+\gamma 
	    \end{array} \right)\vec{v}_{stream}\right|.
\label{stream} 
\end{equation}
\noindent In this work, we restrict our attention to the investigation of the relationship between the microlensing rate due to proper motions and galactic transverse velocity. Note however that the results can easily be converted to describe the corresponding relationship with a stream motion.

\section{the method of comparison}
\label{comp}
We investigate the contribution of stellar proper motions to microlensing statistics by looking for the transverse velocity ($equivalent$ $transverse$ $velocity$) that produces a derivative histogram closest to the one produced by the proper motions of the point masses in the same fields. The similarity of two cumulative distributions $P_{1}$ and $P_{2}$ is quantified by their KS difference D where
\begin{equation}
D=max(|P_{1}-P_{2}|).
\end{equation}
This procedure leads to the natural definition of a constant, which following Kundic, Witt \& Chang (1993), and Wambsganss \& Kundic (1993) we call the effectiveness parameter 
\begin{equation}
a_{tot}\equiv \frac{equivalent\,\, transverse \,\,velocity}{velocity\,\, dispersion}.
\label{a}
\end{equation}
Note that this value differs to that defined by Wambsganss \& Kundic (1993), as it takes account of all microlensing as opposed to only HMEs. For the sake of clarity we therefore refer to the quantity defined by Eqn \ref{a} as the total effectiveness parameter ($a_{tot}$).

Figure \ref{imlc} shows the images of a source line as well as the corresponding point source light curve for a simulation having an optical depth of $\kappa=0.4$. The source trajectory is aligned at 45 degrees to the $x_{1}$-axis.
 Figure \ref{dertest} displays the cumulative histogram of derivatives that was calculated directly from the combination of this light curve with a transverse velocity (dot-dashed line). 
 Each of the point masses in this simulation were given a velocity equal to the transverse velocity and parallel to the source line. In the absence of a shear, the transverse and stream motions are equivalent.
 The distribution of derivatives at stationary source points along the source line was then calculated using the expression \ref{der}, and is shown as the solid light line in Figure \ref{dertest}. In addition, the histogram of derivatives in the transverse motion case was also calculated using the combination of the analytical expression for the light curve derivative (Eqn \ref{der_stat}), with the transverse velocity.

 The histograms calculated from the analytical derivatives agree to within the numerical resolution of the computer (as they must if our expressions are correct). The histograms calculated from the analytical and numerical derivatives (calculated from a 5-point derivative along the light-curve) display excellent agreement, having a maximum KS difference of 0.008.  Figure \ref{dertest} shows the simple nature of the derivative distribution, even when computed over a relatively short sample length. This is in stark contrast to the highly disordered parent light curve.

 As a further check of our method, the stars are again given parallel motions resulting in a stream velocity, but this time the direction of this motion is perpendicular to that of the transverse velocity. In this case our simulation consists of 100 fields of point masses, each with an optical depth of 0.4. The best fit was obtained between the histogram of derivatives due to stream motions, and that due to a transverse velocity, by finding the equivalent transverse velocity that minimised the KS difference between the histograms. This procedure produced a ratio between the value of the stellar stream motion and the calculated equivalent transverse motion of 0.9993 with a minimised KS difference of 0.007. These simulations demonstrate that the method employed is working correctly.

\section{Variation with Optical Depth}
\label{opt_sect}

\subsection{The effect of optical depth}
In order to investigate the dependence of the relative effects of a transverse motion and that of a collection of random proper motions, we start by following the example of Kundic, Witt \& Chang (1993) who explore the unrealistic situation in which the point masses are all solar mass objects that have a motion of fixed magnitude, but a random direction lying in the lens plane. We label simulations with this form for the stellar velocity distribution by $CONST.DISP.$ These models allow us to explore the relationship between the microlensing properties of the two types of motion in a simple situation without the added complexities of velocity and mass functions. These are discussed in sections \ref{mass_func_sect} and \ref{vel_disp_sect}.

\begin{table*}
\caption{\label{opt_dept_val}Results from simulations of sections \ref{opt_sect}, \ref{mass_func_sect} and \ref{vel_disp_sect}. The simulations had no applied shear ($\gamma=0$). Details are found in the text.}
\begin{tabular}{|c|c|c|c|c|c|c|c|c|}
\hline
       &  &                       & Effectiveness         &                       &  &                   & Minimised KS      &                     \\
       &  &                       & Parameter ($a_{tot}$) &                       &  &                   & Difference        &                     \\\\
Optical&  &  $MSOLAR$,            & $SALPETER$,           &   $SALPETER$,         &  &  $MSOLAR$,        & $SALPETER$,       &   $SALPETER$,       \\
Depth  &  & $CONST.$ $DISP.$       & $CONST.$ $DISP.$     &   $GAUSS. DISP.$      &  &  $CONST.$ $DISP.$ & $CONST.$ $DISP.$  &   $GAUSS. DISP.$    \\ \hline \hline  
 0.025 &  & 0.95$\pm$ .07         &   -                   &     -                 &  & 0.011 (.000015)   & -                 &           -           \\
 0.050 &  & 0.87$\pm$ .04         & -                     &           -           &  & 0.010 (.000078)   &   -               &     -                 \\
 0.10  &  & 0.91$\pm$ .02         & -                     &           -           &  & 0.009 (.00067)    &   -               &     -                 \\
 0.20  &  & 0.99$\pm$ .03         & 1.01$\pm$.03          & 1.20$\pm$0.01         &  & 0.010 (.0071)     & 0.007 (.014)      &  0.010 (.013)          \\ 
 0.30  &  & 1.08$\pm$ .02         & 1.10$\pm$.03          & 1.28$\pm$0.03         &  & 0.012 (.014)      & 0.009 (.028)      &  0.010 (.037)          \\
 0.40  &  & 1.20$\pm$ .03         & 1.17$\pm$.02          & 1.46$\pm$0.04         &  & 0.018 (.083)      & 0.008 (.049)      &  0.010 (.050)          \\ 
 0.50  &  & 1.32$\pm$ .07         & 1.31$\pm$.06          & 1.60$\pm$0.03         &  & 0.009 (.131)      & 0.007 (.088)      &  0.004 (.076)          \\ 
 0.60  &  & 1.38$\pm$ .05         & 1.39$\pm$.06          & 1.77$\pm$0.04         &  & 0.007 (.132)      & 0.006 (.130)      &  0.003 (.127)          \\ \hline
 
\end{tabular}

\end{table*}

\begin{figure}
\vspace*{55mm}
\centerline {
\includegraphics{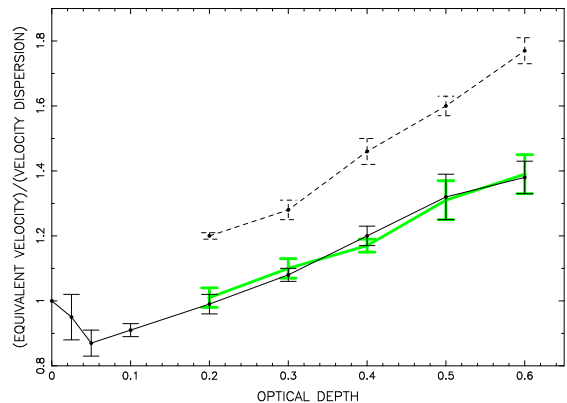}
}
\caption{ The relationship between optical depth $\kappa$ and the effectiveness parameter $a_{tot}$. In these simulations there was no applied shear ($\gamma=0$). The thin, thick and dashed lines show the dependence for simulations of types $MSOLAR/CONST.DISP.$, $SALPETER/CONST.DISP.$ and $SALPETER/GAUSS.DISP.$ respectively.}
\label{optical_gr}
\end{figure}

We consider the effect of optical depth on the value of $a_{tot}$. $MSOLAR$ type simulations were made at optical depths between $\kappa=0.025$ and $\kappa=0.6$ (simulations 1-8 in Table \ref{parameters}). Table \ref{opt_dept_val} displays the value of $a_{tot}$ obtained as well as the best fit KS difference, and the derivative at which the KS difference was found (in brackets). Table \ref{opt_dept_val} also displays the results of sections \ref{mass_func_sect} and \ref{vel_disp_sect}. The quoted error in these, and all values of $a_{tot}$ presented in this paper have been estimated by dividing the total simulation at each optical depth into 5 smaller simulations, and calculating the standard deviation in the resulting values.

Figure \ref{optical_gr} shows graphically the relationship between $\kappa$ and $a_{tot}$. The results of this section are represented by the thin dark line on this plot. We find that at high optical depth the effect of random proper motions on the microlensing rate is larger than that of a transverse velocity of the same magnitude. However at low optical depth the proper motions produce a lower microlensing rate. The total effectiveness parameter, or any similar quantity must have a value of 1 in the limit of zero optical depth, and we find evidence that there is a trend towards a total effectiveness parameter of 1 in this limit. This has not been pursued further, because in the case of Q2237+0305 we are interested in optical depths of between about $\kappa=0.2$ and $\kappa=0.6$.

 \begin{table}
\begin{center}
\caption{\label{tab_HAE}Results from simulations of type $MSOLAR$, with velocity dispersions that are of type $CONST.DISP$. The simulations had no applied shear ($\gamma=0$). Details are found in the text.}
\begin{tabular}{|c|c|c|}
\hline
Optical  & Effectiveness Parameter & Effectiveness Parameter \\
Depth    &  $a_{HME}$ ($\frac{d\,\Delta m}{d\,t}>0.15$)  &  $a_{HME}$ ($\frac{d\,\Delta m}{d\,t}>1 $) \\ \hline\hline
 0.10    &   1.02                       &     1.01                       \\
 0.20    &   1.09                       &     1.08                       \\
 0.30    &   1.16                       &     1.15                       \\
 0.40    &   1.20                       &     1.13                       \\
 0.50    &   1.30                       &     1.22                       \\
 0.60    &   1.34                       &     1.29                       \\ \hline
\end{tabular}

\end{center}
\end{table}
\begin{figure*}
\vspace*{105mm}
\centerline {
\includegraphics{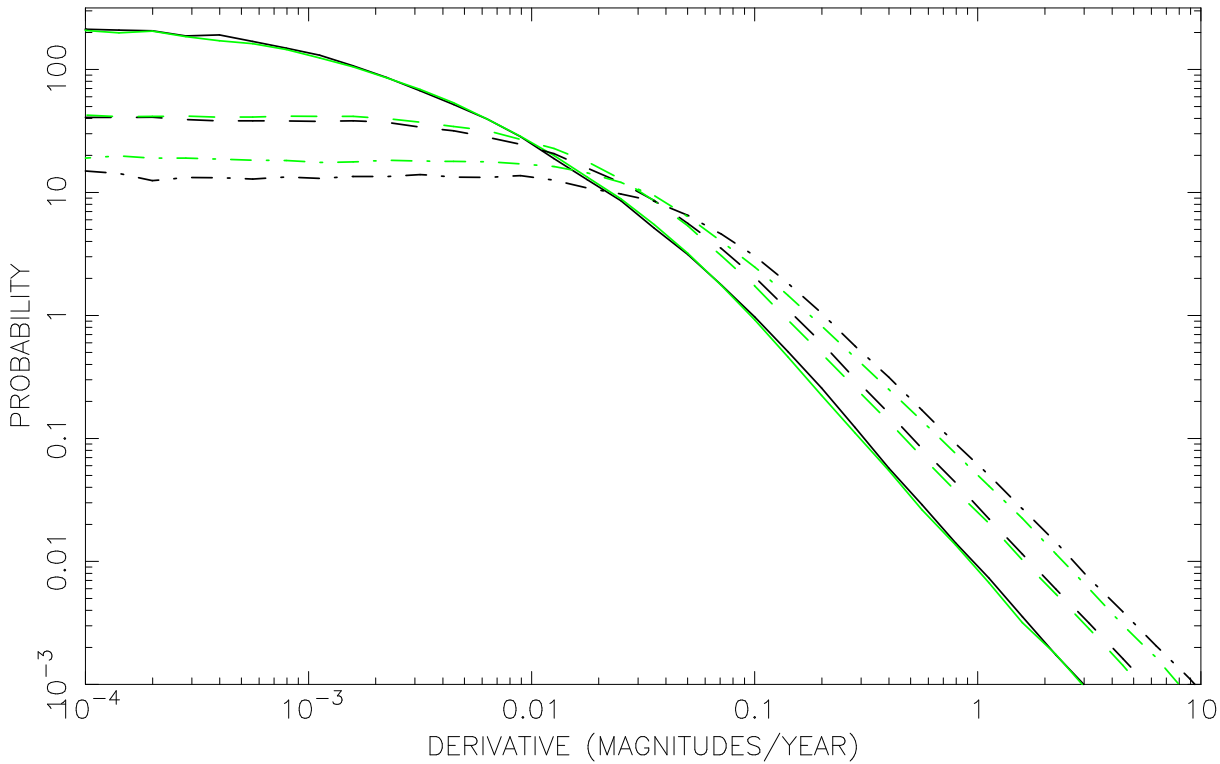}
}
\caption{The probability density functions in both the cases of microlensing due to a transverse velocity (light lines), and due to stellar proper motions (dark lines). The probability density functions are shown for the cases of $\kappa=0.2$ (solid lines), $\kappa=0.4$ (dashed lines), and $\kappa=0.6$ (dot-dashed lines). The simulations had no applied shear ($\gamma=0$).}
\label{prob_dense}
\end{figure*}

The reasons for the behaviour of $a_{tot}$ at both low and high optical depths are illuminated through the comparison of probability density functions of light curve derivatives produced both by stellar proper and galactic transverse motions at various optical depths. Figure \ref{prob_dense} displays these probability density functions for $\kappa=0.2$, $0.4$ and $0.6$ in the case where both the magnitudes of the proper motions and of the transverse velocity are $300\,km\,sec^{-1}$. In this diagram the dark lines indicate the probability density function for light curve derivatives produced through proper motion, while the light lines indicate the function for a static field with a transverse velocity. All three cases display an excess probability of finding a large light curve derivative in the case of microlensing due to the proper motion of stars. The increase in microlensing rate at higher optical depth is also illustrated by these functions.

 The excess of large derivatives in light curves that result from proper motions is quantified in Table \ref{tab_HAE}. In this case the HME effectiveness parameter $a_{HAE}$ is defined as the ratio between the speed of the point masses, and that of the transverse velocity required to produce the same probability of finding a light curve derivative above some minimum level. The values we have chosen for this minimum level are $\frac{d\,\Delta m}{d\,t}=0.15$, which corresponds to the value used by Wambsganss \& Kundic (1993) scaled by a transverse velocity of 300$\,km\,sec^{-1}$, and $ \frac{d\,\Delta m}{d\,t}=1$. This measure differs from that employed by Wambsganss \& Kundic (1993) however because we do not require a minimum magnification shift, but rather just a minimum rate of change. If these large derivatives are only found during HMEs, then $a_{HME}$ approximates the ratio of probabilities between the two cases that the source is undergoing an HME at any given time. This is not necessarily the same thing as the ratio of the number of HMEs produced by the two types of motion because it does not take into account the typical HME time-scale. The results presented in Table \ref{tab_HAE} show that the trend of an increasing effectiveness parameter with optical depth is present in the high derivative regime, consistent with the results of Kundic, Witt \& Chang (1993). The excess of large derivatives is expected following the calculations of Wambsganss \& Kundic (1995) who show that events produced by random proper motions are on average shorter, more frequent and have larger slopes. 

Figure \ref{prob_dense} shows a change in behaviour in the low derivative regime. At the higher optical depths considered, the increase in large derivatives is offset by a corresponding decrease in the probability of finding a low derivative. However, at the lower end of the range there is an increase both in the probability of finding the highest, and the lowest derivatives in the proper motion case. This increased population of small derivatives dominates the evaluation of $a_{tot}$ at low optical depth because the lowest derivatives are by far the most common. This explains why the total effectiveness parameter dips below 1 at low optical depth. Figure \ref{optical_gr} suggests that this effect is maximised at an optical depth of around $\kappa=0.05$, while at optical depths approaching zero the point masses should behave independently (in terms of microlensing), producing an effectiveness parameter of $a_{tot}=1$.

 As the optical depth is increased, the population of caustics becomes more dense. This increase in density corresponds to an increase in the proportion of large light curve derivatives, as well as a decrease in the proportion of low derivatives. The increase in the contribution of fold caustics is reflected in both the increased effectiveness parameter, as well as the location of the derivative where the minimum KS difference is found. The relatively small size of this derivative highlights the fact that the total effectiveness parameter is a representation of microlensing flux variations of all levels.

\subsection{The effect of a mass function}
\label{mass_func_sect}

\begin{figure}
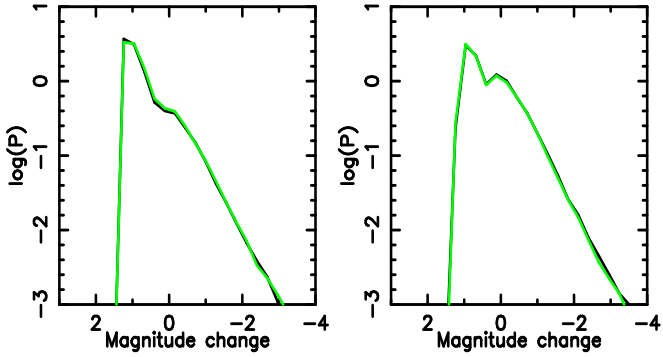

\vspace*{55mm}
\includegraphics{fig5a.epsi}
\includegraphics{fig5b.epsi}
\caption{Magnification distributions. Left: $\kappa=0.3$, $\gamma=0$, $MSOLAR$ (light line) and $SALPETER$ (dark line). Right: $\kappa=0.3$, $\gamma=+0.4$ (light line) and $\gamma=-0.4$ (dark line).}
\label{amp_dist}
\end{figure}

\begin{figure*}
\vspace*{105mm}
\centerline {
\includegraphics{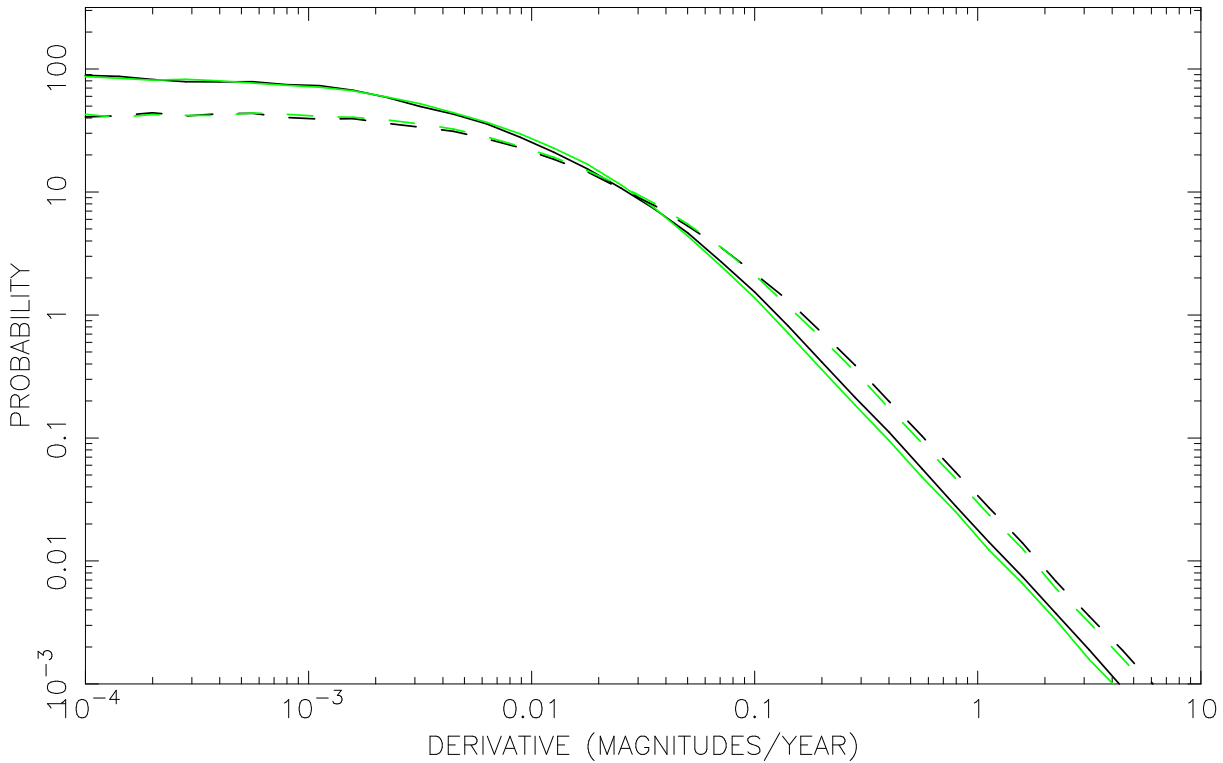}
}
\caption{The probability density functions in both the cases of microlensing due to a transverse velocity (light lines), and due to stellar proper motions (dark lines). The probability density functions are shown for simulations of type $MSOLAR$ (solid lines) and $SALPETER$ (dashed lines). The optical depth in these simulations was $\kappa=0.3$, there was no applied shear ($\gamma=0$). }
\label{prob_dense_mass_func}
\end{figure*}

Previous work has demonstrated that the characteristic time-scale for variability scales as $\sqrt{\langle m \rangle}$, but is independent of the form of the microlens mass function (Witt, Kaiser \& Refsdal 1993; Lewis \& Irwin 1996). Also, the probability distribution of magnifications is a function only of the parameters of optical depth and shear (eg. Lewis \& Irwin 1995). This is demonstrated for the case of models 5 and 10 in the left hand plot of Figure \ref{amp_dist}. We therefore expect that the distribution of microlensed light curve derivatives is a function of $\sqrt{\langle m \rangle}$, but that $a_{tot}$ is independent of the mass function.  

We investigate this by repeating some of the simulations of section \ref{opt_sect} with a microlensing model of type $SALPETER$ (models 9-13 in Table \ref{parameters}). With the introduction of the mass function the mean mass of stars in the model has been lowered. There is therefore an increase in the number density of stars in this model, and a corresponding increase in the number density of caustics. The increase in caustic density at constant optical depth is not expected to have an effect on $a_{tot}$ because the rates of microlensing due to stellar proper motions and transverse velocity are increased equally. Figure \ref{prob_dense_mass_func} demonstrates this behaviour through comparison of derivative histograms that result from $MSOLAR$ type simulations (solid lines) and $SALPETER$ type simulations (dashed lines). The histograms of proper motion light curve derivatives are represented by the dark lines while the transverse motion histograms are represented by the light lines. The transverse velocities and the magnitudes of the proper motions that produce the microlensing in these histograms are each $300\,km\,sec^{-1}$. The histograms constructed from the $SALPETER$ model show an increase in the number of large derivatives and a decrease in the number of small derivatives, demonstrating the increase in microlensing rate where the mean mass of stars is lower. The effect is applicable equally to both the proper motion histogram, and the transverse velocity histogram. This is demonstrated by the values of $a_{tot}$ in Table \ref{opt_dept_val}, which can be compared with those obtained in section \ref{opt_sect}. These values are also plotted in Figure \ref{optical_gr} (thick light line). The results demonstrate that $a_{tot}$ is independent of the mass function.

\subsection{The Effect of a Velocity dispersion}
\label{vel_disp_sect}

Sections \ref{opt_sect} and \ref{mass_func_sect} have assumed the unrealistic, but easily interpreted case where all stars are given an equal speed in a random direction in the lens plane. These models produce distributions for light curve derivatives that result from the proper motion of stars, and from the equivalent transverse velocity that have KS differences of $\approx 10^{-2}$. We need to check that this level of equivalence holds in the more realistic case where the stars are given a dispersion of velocities rather than a constant value. For models in this section, the components of the stellar velocities were each assigned at random according to a Gaussian distribution. We label this type of distribution by $GAUSS.DISP$. Simulations of type $SALPETER$ were made at various optical depths which cover the range of interest in the case of Q2237+0305.

 With the inclusion of a velocity dispersion, $a_{tot}$ has been redefined as the equivalent transverse velocity divided by the average stellar speed. Table \ref{opt_dept_val} shows the dependence of $a_{tot}$ on $\kappa$, as well as the minimised KS-difference and the derivative at which it is found (in brackets). The trend, which is shown in Figure \ref{optical_gr} (dashed line) is similar to the one obtained at the same optical depths in sections \ref{opt_sect} and \ref{mass_func_sect}. However the value of the total effectiveness parameter is larger at each optical depth considered. Kundic, Witt \& Chang (1993) have shown that a velocity dispersion produces a higher average caustic velocity (with respect to the average stellar speed). The high velocity caustics produce a greater proportion of large light curve derivatives, and consequently leave a dearth of small derivatives. Importantly, the KS difference between the proper motion and scaled transverse histograms is again of order $10^{-2}$ in these models. The method outlined here can therefore be used to analyse the effect of proper motions in realistic microlensing simulations of objects such as Q2237+0305.

\section{Variation with Shear}
\label{shear_sect}

\begin{table}
\begin{center}
\caption{\label{tab_shear}Results from simulations of type $MSOLAR$, with velocity dispersions that are of type $CONST.DISP.$ The optical depth in these simulations was $\kappa=0.3$. Details are found in the text.}
\begin{tabular}{|c|c|c|}
\hline
Shear   & Effectiveness          & KS                \\ 
        & Parameter ($a_{tot}$)  & Difference        \\ \hline \hline
 -0.4   & 1.42$\pm$.05           & 0.006 (.03)       \\
 -0.2   & 1.28$\pm$.04           & 0.005 (.014)     \\
 0.0    & 1.08$\pm$.02           & 0.012 (.007)     \\
 0.20   & 0.97$\pm$.02           & 0.016 (.017)     \\
 0.40   & 0.92$\pm$.02           & 0.009 (.029)     \\ \hline

\end{tabular}
\end{center}
\end{table}

\begin{figure}
\vspace*{60mm}
\centerline {
\includegraphics{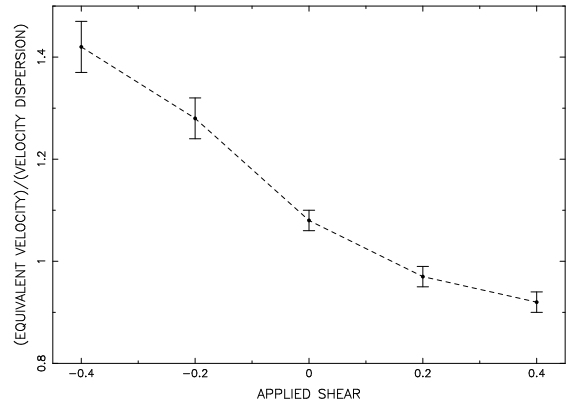}
}
\caption{The relationship between the applied shear $\gamma$ and the effectiveness parameter $a_{tot}$. In these simulations, the optical depth was $\kappa=0.3$.}
\label{shear_gr}
\end{figure}

\begin{figure*}
\vspace*{105mm}
\centerline {
\includegraphics{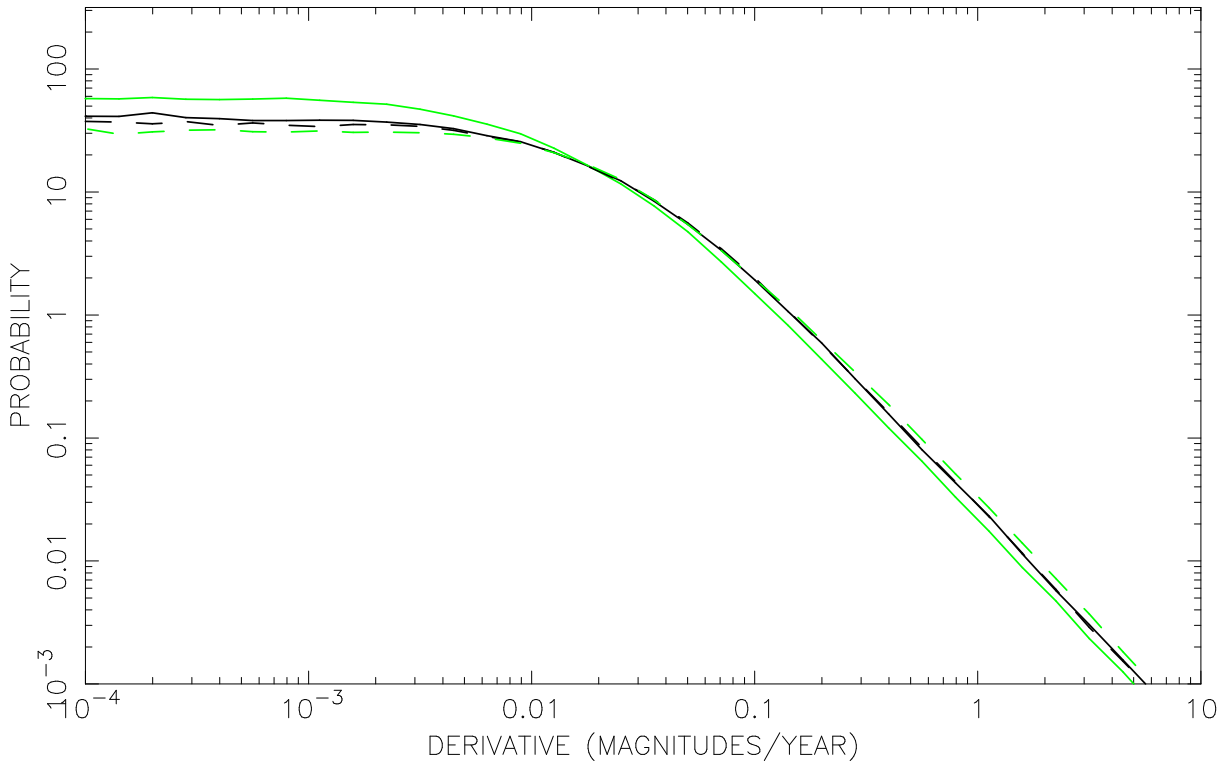}
}
\caption{The probability density functions in both the cases of microlensing due to a transverse velocity (light lines), and due to stellar proper motions (dark lines). The probability density functions are shown for the cases of $\gamma=-0.4$ (solid lines), and $\gamma=+0.4$ (dashed lines). The optical depth in these simulations was $\kappa=0.3$. }
\label{prob_dense_shear}
\end{figure*}

The addition of an applied shear on the model starfield breaks the isotropy of the model. Kundic, Witt \& Chang (1993) have shown that in the presence of an applied shear, the mean caustic velocity is dependent on the direction along which it is measured. In the case of microlensed flux variation of a stationary source that results from stellar proper motions, the only component of the model that has a preferential direction is the shear vector. The rate of microlensing produced through proper motions (by any measure) is therefore not dependent on the direction in the source plane along which the light curve derivatives are sampled. The magnification distribution is also independent of the orientation of the shear (eg. Lewis \& Irwin 1995), an independence that is demonstrated for models 14 and 17 in the right hand plot of Figure \ref{amp_dist}. However the rate of microlensing from the transverse motion of the source with respect to a static starfield that is perturbed by an applied shear is dependent on the relative orientations of the transverse velocity and the shear (eg. Witt, Kaiser \& Refsdal 1993; Lewis \& Irwin 1996). The effect of an applied shear on the caustic network is to stretch the caustics along a direction that is at right angles to the shear ($\gamma<0$). The result of this directional clustering is that a source moving in a direction that is parallel to the shear ($\gamma>0$) experiences a higher microlensing rate. The directional dependence of the microlensing rate results in the effectiveness parameter also being dependent on the direction of the transverse motion.

Table \ref{tab_shear} shows the total effectiveness parameter at various values of applied shear. The values were computed for a field having an optical depth of $\kappa=0.3$ in $1M_{\odot}$ stars ($MSOLAR$). The stars in these models were given a proper motion of type $CONST.DISP.$ The dependence of $a_{tot}$ on $\gamma$ is plotted in Figure \ref{shear_gr}. This plot shows that the addition of a shear that is oriented at right angles to the source trajectory increases $a_{tot}$, while a parallel shear reduces $a_{tot}$ with respect to that of the same field in the absence of an applied shear. Table \ref{tab_shear} also shows the minimised KS difference and the derivative at which it is found (in brackets). With the addition of an applied shear, the distributions of light curve derivatives resulting from proper motions, and from a static field in combination with the equivalent transverse velocity have best fit KS differences that are again of order $10^{-2}$. Table \ref{tab_H_shear} shows the variation of $a_{HME}$ with shear. As in the case of variation with optical depth, $a_{HME}$ follows a similar trend to that of $a_{tot}$. 

Figure \ref{prob_dense_shear} displays probability density functions of light curve derivatives produced both by a transverse velocity (light lines) and proper motions (dark lines) (each having a magnitude of $300\,km\,sec$) in the cases of $\gamma=-0.4$ (solid lines) and $\gamma=+0.4$ (dashed lines). As required the distribution produced by proper motions is independent of the sign and therefore direction of the shear. However the transverse velocity produces a histogram of light curve derivatives that is dependent on the relative orientation of the source trajectory and shear. In the case where the shear is aligned at right angles to the source ($\gamma<0$), lower derivatives are more common, and higher derivatives less common than in the case where the shear is aligned with the source trajectory ($\gamma>0$).

\begin{table}
\begin{center}
\caption{\label{tab_H_shear}Results from simulations of type $MSOLAR$, with velocity dispersions that are of type $CONST.DISP$. The optical depth in these simulations was $\kappa=0.3$. Details are found in the text.}
\begin{tabular}{|c|c|c|}
\hline
Shear  & Effectiveness Parameter & Effectiveness Parameter \\
($\kappa=0.3$)       & $a_{HME}$ ($\frac{d\,\Delta m}{d\,t}>0.15$)  & $a_{HME}$ ($\frac{d\,\Delta m}{d\,t}>1 $) \\ \hline\hline
 -.40    &   1.36                       &     1.30                       \\
 -0.20   &   1.27                       &     1.24                       \\
 0.0    &   1.16                        &     1.01                       \\ 
 0.20    &   1.02                       &     0.98                       \\
 0.40    &   0.84                       &     0.82                       \\ \hline
 \end{tabular}
\end{center}
\end{table}

\section{application to Q2237+0305}
\label{2237_sect}

\begin{table}
\begin{center}
\caption{\label{2237_tab} Values of the total effectiveness parameter for microlensing models of type $SALPETER$ corresponding to the images of Q2237+0305. The models include a velocity dispersion of type $GAUSS.DISP.$}
\begin{tabular}{|c|c|c|c|}
\hline
Image 			   &                       & $a_{tot}$               &                     \\
($\kappa,|\vec{\gamma}|$)  &    0 degrees          & 45 degrees              &  90 degrees         \\ \hline\hline
A,B (0.36,0.40)    	   &    0.99$\pm$.08       &  1.18$\pm$.10           & 1.57$\pm$.10         \\
C (0.69,0.71)              &    1.57$\pm$.05       &  1.71$\pm$.07           & 2.26$\pm$.27         \\
D (0.59,0.61)   	   &    1.26$\pm$.05       &  1.49$\pm$.12           & 1.92$\pm$.15         \\ \hline

 \end{tabular}
\end{center}
\end{table}

\begin{figure}
\vspace*{55mm}
\centerline {
\includegraphics{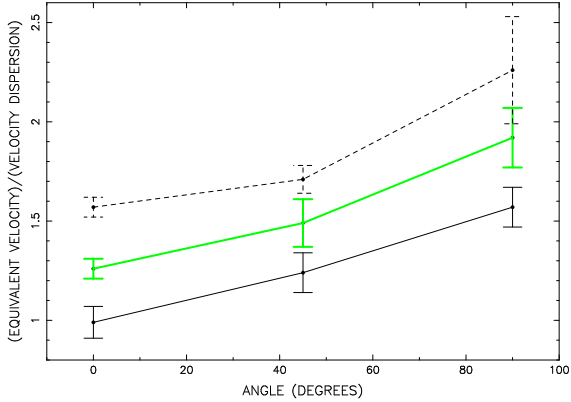}
}
\caption{ The relationship between the effectiveness parameter $a_{tot}$ and the angle between the transverse velocity and shear for the microlensing parameters corresponding to the images in Q2237+0305. Images A/B, C and D are represented by the thin dark, dashed and thick light lines respectively.}
\label{2237_fig}
\end{figure}

As a final consideration we look at the values of $a_{tot}$ corresponding to the parameters that describe the microlensing environment for the images of Q2237+0305. When modelling microlensing in Q2237+0305 we adopt the lensing galaxy model of Schmidt, Webster \& Lewis (1998). This model includes the effect of the bar and produces the following microlensing parameters: image A ($\kappa=0.36$, $|\vec{\gamma}|=0.40$), image B ($\kappa=0.36$, $|\vec{\gamma}|=0.40$), image C ($\kappa=0.69$, $|\vec{\gamma}|=0.71$), image D ($\kappa=0.59$, $|\vec{\gamma}|=0.61$). Simulations of type $SALPETER$ were made for each set of parameters (models 18-26 in Table \ref{parameters}) and $a_{tot}$ found in the cases where the angle between the transverse velocity and the shear is 0, 45 and 90 degrees. The stellar proper motions were modelled by a distribution of type $GAUSS.DISP.$

 The results for $a_{tot}$ are shown in Table \ref{2237_tab}, and plotted in Figure \ref{2237_fig}. As in section \ref{shear_sect} these show that $a_{tot}$ is larger in the case where the transverse velocity is perpendicular to the source ($\gamma<0$). The microlensing parameters ($\kappa$ and $\gamma$) increase in the order of images A/B, D and C. Figure \ref{2237_fig} therefore outlines the general trend of $a_{tot}$ which increases with optical depth and with a shear that is applied at right angles to the source trajectory. The results also demonstrate that for a given set of microlensing parameters, values of $a_{tot}$ vary between the two extreme values for source trajectories that are oriented with respect to the shear at angles other than 0 or 90 degrees. In the case of Q2237+0305, the four images are positioned approximately orthogonally with respect to the galactic centre. Therefore, the direction of galactic transverse motion with respect to the shear is approximately orthogonal in images A and B to that in images C and D. An important point to be noted from Figure \ref{2237_fig} is that when all images of Q2237+0305 are considered together, the relative rates of microlensing due to the transverse motion and stellar proper motions may be dependent on the direction of the galactic transverse motion.

\section{Conclusion}
We have combined the contouring algorithm of Lewis et al. (1993) with analytical expressions for both the derivative of the image magnification resulting from the proper motion of stars, and from the relative motion due to a transverse velocity of the galaxy. This combination has been used to construct probability density functions for point source light curve derivatives in both cases.
 We have compared the two classes of distributions by scaling the derivative in the transverse case to a value that minimises the KS difference. We find that the distribution of light curve derivatives where microlensing is due to proper motion of stars is approximately equivalent to that of microlensing from a static lens with an appropriately scaled transverse velocity. The minimised KS differences are of order $10^{-2}$. The ratio of this appropriately scaled transverse velocity to the mean speed of the stellar proper motions is defined as the total effectiveness parameter $a_{tot}$, and is a measure of the relative rates of microlensing in the two cases. The value of $a_{tot}$ is dependent on the microlensing parameters ($\kappa$ and $\gamma$), but is independent of the assumed microlens mass function. We find that $a_{tot}$ increases with optical depth in the absence of a shear, increases with shear that is aligned at right angles to the transverse source trajectory, and decreases with a shear that is aligned along the transverse source trajectory. The latter results are particularly important in the case of Q2237+0305 where 4 images are placed orthogonally with respect to the centre of the lensing galaxy and so have varying directions of galactic shear with respect to the transverse motion. It may therefore be possible to constrain the direction of transverse galactic motion (assuming $\kappa$ and $\gamma$) from the observed microlensing rate.

\section{acknowledgements}
JSBW acknowledges the support of an Australian Postgraduate Award and a Melbourne University Postgraduate Overseas Research Experience Award. This work was partly supported by NSF grant AST98-02802.

\onecolumn
\begin{appendix}
\section{Expressions For the Derivatives}
\label{app1}

We have defined the following quantities for use in the evaluation of $\frac{d\,\mu_{p}}{d\,t})_{proper}$ and $\frac{d\,\mu_{p}}{d\,t})_{tran}$. 
\begin{eqnarray}
\nonumber
A^{j}&\equiv&m^{j}\frac{(x_{2}-x_{2}^{j})^{2}-(x_{1}-x_{1}^{j})^{2}}{((x_{2}-x_{2}^{j})^{2}+(x_{1}-x_{1}^{j})^{2})^{2}} \\ \nonumber
B^{j}&\equiv& 2m^{j}\frac{(x_{2}-x_{2}^{j})(x_{1}-x_{1}^{j})}{((x_{2}-x_{2}^{j})^{2}+(x_{1}-x_{1}^{j})^{2})^{2}} \\ \nonumber
C^{j}&\equiv&2m^{j}(x_{1}-x_{1}^{j})\frac{3(x_{2}-x_{2}^{j})^{2}-(x_{1}-x_{1}^{j})^{2}}{((x_{2}-x_{2}^{j})^{2}+(x_{1}-x_{1}^{j})^{2})^{3}} \\ \nonumber
D^{j}&\equiv& 2m^{j}(x_{2}-x_{2}^{j})\frac{3(x_{1}-x_{1}^{j})^{2}-(x_{2}-x_{2}^{j})^{2}}{((x_{1}-x_{1}^{j})^{2}+(x_{2}-x_{2}^{j})^{2})^{3}} \\ \nonumber
A&\equiv&\sum_{j=0}^{N_{*}}A^{j}\,\,\,\,\,\,B\equiv\sum_{j=0}^{N_{*}}B^{j} \\
C&\equiv&\sum_{j=0}^{N_{*}}C^{j}\,\,\,\,\,\,D\equiv\sum_{j=0}^{N_{*}}D^{j}
\label{defs}
\vspace{5mm}
\end{eqnarray}
In terms of these quantities, the full expression for the derivative $\frac{d\,\mu_{p}}{d\,t})_{proper}$ is:  
\begin{eqnarray}
\nonumber
\left.\frac{d\,\mu_{p}}{d\,t}\right)_{proper}&=&-2\mu_{p}^{2}\left(\frac{C(\gamma+A)+BD}{B^{2}-(1-(\gamma+A)^{2})}\right.\times     \left.\sum_{j=0}^{N_{*}}\left(\left(\left(1+\left(\gamma+A\right)\right)A^{j}+BB^{j}\right)v_{1}^{j}+\left(BA^{j}-\left(1+(\gamma+A)\right)B^{j}\right)v_{2}^{j}\right)\right)\\ \nonumber
& &+2\mu_{p}^{2}\left(\frac{D(\gamma+A)-BC}{B^{2}-(1-(\gamma+A)^{2})}\right.\times   \left.\sum_{j=0}^{N_{*}}\left(\left(\left(1-(\gamma+A)\right)B^{j}+BA^{j}\right)v_{1}^{j}+\left(BB^{j}-\left(1-(\gamma+A)\right)A^{j}\right)v_{2}^{j}\right)\right)\\ \nonumber
& &+2\mu_{p}^{2}\sum_{j=0}^{N_{*}}\left(C^{j}(\gamma+A)+BD^{j}\right)v_{1}^{j}\\
& &-2\mu_{p}^{2}\sum_{j=0}^{N_{*}}\left(D^{j}(\gamma+A)-BC^{j}\right)v_{2}^{j}
\label{der}
\vspace{5mm}
\end{eqnarray}
The definitions \ref{defs} can also be used in the expression for the static field light curve derivative:
\begin{eqnarray}
\left.\frac{d\,\mu_{p}}{d\,t}\right)_{tran}&=&2\mu_{p}^{2} \left(\left(D(\gamma+A)-BC\right)\times\frac{B-(1-(\gamma+A)m)}{B^{2}-(1-(\gamma+A)^{2})}+ \right.\left.\left(C(\gamma+A)-BD\right)\times\frac{mB-(1+(\gamma+A))}{B^{2}-(1-(\gamma+A)^{2})}\right)\times v_{1_{tran}}      
\label{der_stat}
\end{eqnarray}

\end{appendix}
\twocolumn

\label{lastpage}

\end{document}